\documentclass{appolb}
\usepackage{graphicx}

\begin{document}
\title{Gamow Shell Model description of $^7$Li and elastic scattering reaction $^4$He($^3$H, $^3$H)$^4$He%
\thanks{Presented at Zakopane 2022 Conference on Nuclear Physics}%
}
\author{J.P. Linares Fernandez, M. P{\l}oszajczak
\address{Grand Acc\'el\'erateur National d'Ions Lourds (GANIL), CEA/DSM - CNRS/IN2P3, BP 55027, F-14000 Caen, France}
\\[3mm]
{N. Michel 
\address{Institute of Modern Physics, Chinese Academy of Sciences, Lanzhou 730000, China}
}
}
\maketitle
\begin{abstract}
Spectrum of $^7$Li and elastic scattering reaction $^4$He($^3$H, $^3$H)$^4$He are studied using the unified description of the Gamow shell model in the coupled-channel formulation (GSMCC). The reaction channels are constructed using the cluster expansion with the two mass partitions [$^4$He + $^3$H], [$^6$Li + n]. 
\end{abstract}
  
\section{Introduction}
 The properties of radioactive nuclei are strongly affected by couplings to many-body continuum of scattering and decay channels.  This calls for a unified theory of these nuclei which would involve a comprehensive description of bound states, resonances and scattering many-body states within a single theoretical framework. A pioneering attempts in this direction were done within the continuum shell model \cite{Okolowicz2003,bennaceur_2000,rotureau_2006,volya_2005}, and has been extended to \textit{ab initio} description of structure and reactions of light nuclei within the no-core shell model coupled with the resonating-group method  \cite{quaglioni_2008} and the no-core shell model with continuum \cite{baroni_2013a}.

An alternative approach within a unifying framework has been proposed with the open quantum system formulation of the shell model, the Gamow Shell Model (GSM) \cite{michel_review_2009,michel_book_2021}. GSM offers the most general treatment of couplings between discrete and scattering states, using Slater determinants defined in the Berggren ensemble of single-particle states. For the description of scattering properties and reactions, GSM should be formulated in the representation of reaction channels (GSMCC) \cite{jaganathen_2014}.  

\section{Theoretical framework}
\label{Formalism}
{
In this section we will briefly outline the GSMCC formalism for the channels constructed with different mass partitions. Detailed discussion of the GSMCC in a single mass partition case can be found in Refs.~\cite{jaganathen_2014,mercenne_2019,michel_book_2021}.
Here, we will mainly concentrate on the differences between one and many mass partition case of the GSMCC and apply for the description of spectrum of $^7$Li and the elastic scattering reaction $^4$He($^3$H,$^3$H)$^4$He.

The ${ A }$-body state of the system is decomposed into reaction channels defined as binary clusters:
\begin{equation}
   |{ \Psi }_{ M }^{ J } \rangle = \sum_{\rm  c } \int_{ 0 }^{ +\infty } |{ \left( {\rm c} , r \right) }_{ M }^{ J } \rangle \frac{ { u }_{\rm c }^{JM} (r) }{ r } { r }^{ 2 } ~ dr \; ,
  \label{scat_A_body_compound}
\end{equation}
where the radial amplitude ${ {u}_{\rm c }^{JM}(r) }$, describing the relative motion between the two clusters in a channel ${\rm c }$, is the solution to be determined for a given total angular momentum ${J}$ and its projection ${M}$. 
 Different channels in the sum (\ref{scat_A_body_compound}) are orthogonalized independently of the partition of neutrons and protons in various binary clusters. The integration variable ${ r }$ is the relative distance between the center-of-mass (CM) of the clusters, and the binary-cluster channel states are defined as:
\begin{equation}
 | \left( {\rm c} , r \right)\rangle  = \hat{ \mathcal{A}} | \{ | \Psi_{{\rm  T} , J_{ \rm T }; N-n, Z-z} \rangle 
 \otimes  | r ~ \ell ~ J_{\rm int} ~ J_{\rm P}; n, z \rangle \}_{ M }^{J} \rangle  
  \label{channel}
\end{equation} 
The channel index ${\rm c}$ stands for different quantum numbers and mass partitions $\{ (N - n, Z - z, { J }_{ \rm T}) ; (n , z, { L } , J_{{\rm int}}, J_{\rm P}) \}$, where $n$ and $z$ are the number of neutrons and protons of a projectile and $N$ and $Z$ are the total number of neutrons and protons in the combined system of a projectile and a target. ${\hat{ \mathcal{A}}}$ is the inter-cluster antimmetrizer that acts among the nucleons pertaining to different clusters. 
The states $|{\Psi_{{\rm  T} , J_{ \rm T } }}\rangle$ and $|{r ~ \ell ~ J_{{\rm int}} ~ J_{\rm P}}\rangle$ are the target and projectile states in the channel 
$|{ \left( {\rm c} , r \right)}\rangle$ (\ref{channel}) with their associated total angular momentum ${ { J }_{ \rm T } }$ and ${ { J }_{ \rm P } }$, respectively.
The angular momentum couplings read $\mathbf{ J}_{\rm P } = \mathbf{ J}_{ \rm int } + \mathbf{{\ell}}$ and  
${ \mathbf{ J}_{\rm A} = \mathbf{J}_{\rm P} + \mathbf{ J}_{\rm T} } $.

The Schr{\"o}dinger equation $H |{\Psi_{M_{ A }}^{J_{ A }}}\rangle = E |{\Psi_{M_{ A }}^{J_{ A }}}\rangle$ in the channel representation of the GSM takes the form of coupled-channel equations:
\begin{equation}
  \sum_{\rm c}\int_{0}^{\infty}  \!\!\! r^{ 2 } \left( H_{\rm cc' } (r , r') - E N_{\rm cc' } (r , r') \right) \frac{ { u }_{\rm c } (r) }{ r } = 0	\ ,
  \label{cc_cluster_eq}
\end{equation}
where ${ E }$ stands for the scattering energy of the ${ A }$-body system. To simplify reading, we have dropped the total angular momentum labels ${ J }$ and ${ M }$, but one should keep in mind that the resolution of Eq. (\ref{cc_cluster_eq}) is done for fixed values of ${J}$ and ${ M }$.
The kernels in Eq. (\ref{cc_cluster_eq}) are defined as:
\begin{eqnarray}
   H_{\rm cc' } (r,r') &=& \langle{ ({\rm c},r) }| \hat{ H } |{({\rm c'},r') }\rangle \label{h_cc_compound} \\
   N_{\rm cc' } (r,r') &=& \langle{ ({\rm c},r) | ({\rm c'},r') }\rangle \label{n_cc_compound}
\end{eqnarray}

As the nucleons in the target and those of the projectile can no longer interact with each other at high energy due to their large difference of momenta, it is convenient to express the Hamiltonian $\hat{ H }$ as $ \hat{ H } = \hat{ H }_{ \rm T } + \hat{ H }_{ \rm P } + \hat{ H }_{ \rm TP }$, 
where ${ \hat{ H }_{ \rm T } }$ and ${ \hat{ H }_{ \rm P } }$ are the Hamiltonians of the target and projectile, respectively.
In particular, ${ { \hat{ H } }_{ \rm T } }$ is the CM intrinsic Hamiltonian of the target.
The projectile Hamiltonian ${ { \hat{ H } }_{ \rm P } }$ can be decomposed as follows: 
${ { \hat{ H } }_{ \rm P } = { \hat{ H } }_{ {\rm int} } + { \hat{ H } }_{ {\rm CM} } }$,
where ${ { \hat{ H } }_{\rm  int } }$ describes its intrinsic properties of the projectile, and ${ { \hat{ H } }_{ \rm CM } }$ describes the movement of its center of mass. 
Thus, the inter-cluster Hamiltonian ${ { \hat{ H } }_{ \rm TP } }$ is defined as: ${ { \hat{ H } }_{ \rm TP } = \hat{ H } - { \hat{ H } }_{ \rm T } - { \hat{ H } }_{ \rm P } }$, where ${ \hat{ H } }$ is the standard GSM Hamiltonian.

The channel states ${ |{ (c,r) } }\rangle$ are expanded  in a one-body Berggren basis to calculate the kernels $H_{\rm cc' } (r,r')$ and $N_{ \rm cc' } (r,r')$:
\begin{equation}
  | ({\rm c},r) \rangle = \sum_{n} \frac{ { u }_{ n \ell } (r) }{ r } |{ ({\rm c},n) }\rangle \ ,
  \label{expansion_channel_n}
\end{equation}
where $$ |{ ({\rm c},n) }\rangle = \hat{ \mathcal{A}} | \{ |{\Psi_{{ \rm T }, {J_{ \rm T }} }}\rangle \otimes | n ~ \ell ~ J_{{\rm int}} ~ J_{\rm P} \rangle \}_{ M }^{J} \rangle \ ,$$ with 
${ { \hat{ H } }_{\rm CM } |{ n \ell }\rangle = { E }_{\rm  CM } |{ n \ell } }\rangle$, and ${ n }$ refers to the projectile CM shell number in the Berggren basis state. For simplicity, the spin-dependence has beem omitted in the notation of $u_{n\ell}(r)$.
The basis of ${ |{ n \ell }\rangle }$ states is then generated by diagonalizing ${ { \hat{ H } }_{\rm CM } }$, which contains a kinetic term and a one-body potential of Wood-Saxon (WS) type with a spin-orbit term generated by nucleons of the target.

The many-body matrix elements of the norm kernel Eq. (\ref{n_cc_compound}) are calculated using the Slater determinant expansion of the cluster wave functions ${ |{ ({\rm c},n) }\rangle }$.
The treatment of the non-orthogonality of channels is the same as in the one-nucleon projectile case \cite{jaganathen_2014}.
Note that the antisymmetry of channels, enforced by the antisymmetrizer in Eq. (\ref{channel}), is exactly taken into account through the expansion of many-body targets and projectiles with Slater determinants.
Once the kernels are computed, the coupled-channel equations (\ref{cc_cluster_eq}) can be solved using a numerical method based on a Berggren basis expansion of the Green's function ${ { (H - E) }^{ -1 } }$, that takes advantage of GSM complex energies \cite{mercenne_2019,michel_book_2021}. 

The GSMCC Hamiltonian is Hermitian because the matrix elements are calculated in the harmonic oscillator basis. However, the calculation of resonances using this Hamiltonian is done in the Berggren basis, so that the GSMCC Hamiltonian matrix becomes complex symmetric. The cross sections are calculated by coupling the real-energy incoming partial waves to the target states given by the Hermitian Hamiltonian. Consequently, the framework related to cross section calculation is fully Hermitian, whereas complex energies arise for resonances because one diagonalizes the  Hamiltonian matrix in Berggren basis which  is complex symmetric.

\section{Effective Hamiltonian}
\label{Hamiltonian}
The effective Hamiltonian is optimized using the GSM in Berggren basis. The model is formulated in relative variables of the cluster orbital shell model~\cite{Suzuki1988} what allows to eliminate spurious CM excitations. In the following, we shall use $^4$He inert core with two or three valence nucleons to describe $^6$Li or $^7$Li wave functions, respectively. 

The Hamiltonian consists of the one-body part and the nucleon-nucleon interaction of FHT type~\cite{1979Furutani} supplemented by the Coulomb term $V_{\rm FHT} = V_{\rm c} + V_{\rm LS} + V_{\rm T} + V_{\rm Coul}$, 
where $V_{\rm c}$ , $V_{\rm LS}$, $V_{\rm T}$ represent its central, spin-orbit and tensor part, respectively. The two-body Coulomb potential 
$V_{\rm Coul}(r) = e^2/r$ between valence protons is treated exactly by incorporating its long-range part into the basis potential \cite{PRC_isospin_mixing}.
 The $^4$He core is mimicked by a one-body potential of the WS type, with a spin-orbit term, and a Coulomb field. The Coulomb potential is generated by a spherical Gaussian charge distribution: $U_{\rm Coul}(r)=2e^2{\rm erf}(r/{\tilde R}_{\rm ch})/r$, where ${\tilde R} = 4R_{\rm ch} /(3\sqrt{\pi} )$. The parameters of the one-body potential and the FHT interaction have been adjusted to the spectra of $^6$Li, $^7$Be, and $^7$Li. The parameters of the interaction obtained in the present optimization agree with those of Ref.~\cite{PRC_FHT_Yannen} within the statistical errors.

\subsection{Model space in GSMCC calculation}
\label{Model space}
GSM calculations of $^6$Li target are performed in the approximation of $^4$He core and valence particles in two resonant-like shells: $0p_{3/2}$, $0p_{1/2}$, and many shells $\{p_{3/2}\}$, $\{p_{1/2}\}$, $\{s_{1/2}\}$, $\{d_{5/2}\}$, $\{d_{3/2}\}$, $\{f_{7/2}\}$, $\{f_{5/2}\}$  which approximate the non-resonant continuum. The calculation is done in the harmonic oscillator (HO) basis and the continuum is approximated by 5 lowest HO wave functions for each $\ell$ and $j$.
Antisymmetric eigenstates of the GSMCC are expanded in the basis of channels: 
$[{^4}{\rm He}(0^+_1)\otimes{^3}{\rm H}(L_j)]^{J^{\pi}}$ and $[{^6}{\rm Li}(K^{\pi})\otimes{\rm n}(\ell_j)]^{J^{\pi}}$ for $^7$Li.
 Channels with one-nucleon projectile are built by coupling the GSM wave functions for the ground state $K^{\pi}=1^+_1$ and excited states $K^{\pi} = 1^+_2$, $3^+_1$, $0^+_1$, $2^+_1$, $2^+_2$ of $^{6}$Li with the neutron wave functions in several partial waves $\ell_j$: $s_{1/2}$, $p_{1/2}$, $p_{3/2}$, $d_{3/2}$, $d_{5/2}$}, $f_{5/2}$, $f_{7/2}$. 

The internal structure of $^3$H projectile in the channels $[{^4}{\rm He}(0^+_1)\otimes{^3}{\rm H}(L_j)]^{J^{\pi}}$ is calculated using the N$^3$LO interaction~\cite{PRC_N3LO} without the three-body contribution, fitted on phase shifts properties of proton-neutron elastic scattering reactions. The N$^3$LO realistic interaction is diagonalized in six HO shells to generate intrinsic states of $^3$H. The oscillator length in this calculation is $b=1.65$ fm. For this value of the oscillator length, the ground state energy of $^3$H is -7.14 MeV as compared to -8.48 MeV experimentally. In the coupled-channel equations of GSMCC, we use the experimental binding energies of $^3$H to assure a correct $^4$He+$^3$H threshold energy.
The relative motion of the $^3$H cluster CM and the $^4$He target is calculated in the Berggren basis generated  by proton and neutron WS potentials. For the intrinsic $^3$H wave function we consider only the most important one bearing $J^{\pi}_{{\rm int}} = 1/2^+$. Since the CM part of $^3$H projectile has $L_{\rm CM} \leq 3$, therefore the total angular momentum of $^3$H projectile satisfies $J_{\rm tot} \leq 7/2$.

The use of both realistic interaction for a projectile and effective Hamiltonian for composites induces no problem in the GSMCC framework. Before and after the reaction occurs, $^3$H is far from the target and its properties are prominent, whereas during the reaction the properties of  $^7$Li are decisive. As the FHT interaction is defined from $^6$Li, $^7$Be, $^7$Li properties, it cannot grasp the structure of $^3$H at large distances.
Conversely, the N$^3$LO interaction cannot be used in the approximation of a core and valence particles.

\section{Results}
\label{Results}
\begin{figure}[htb]
\vskip -1.2truecm
\begin{center}
\includegraphics[width=8cm,angle=0]{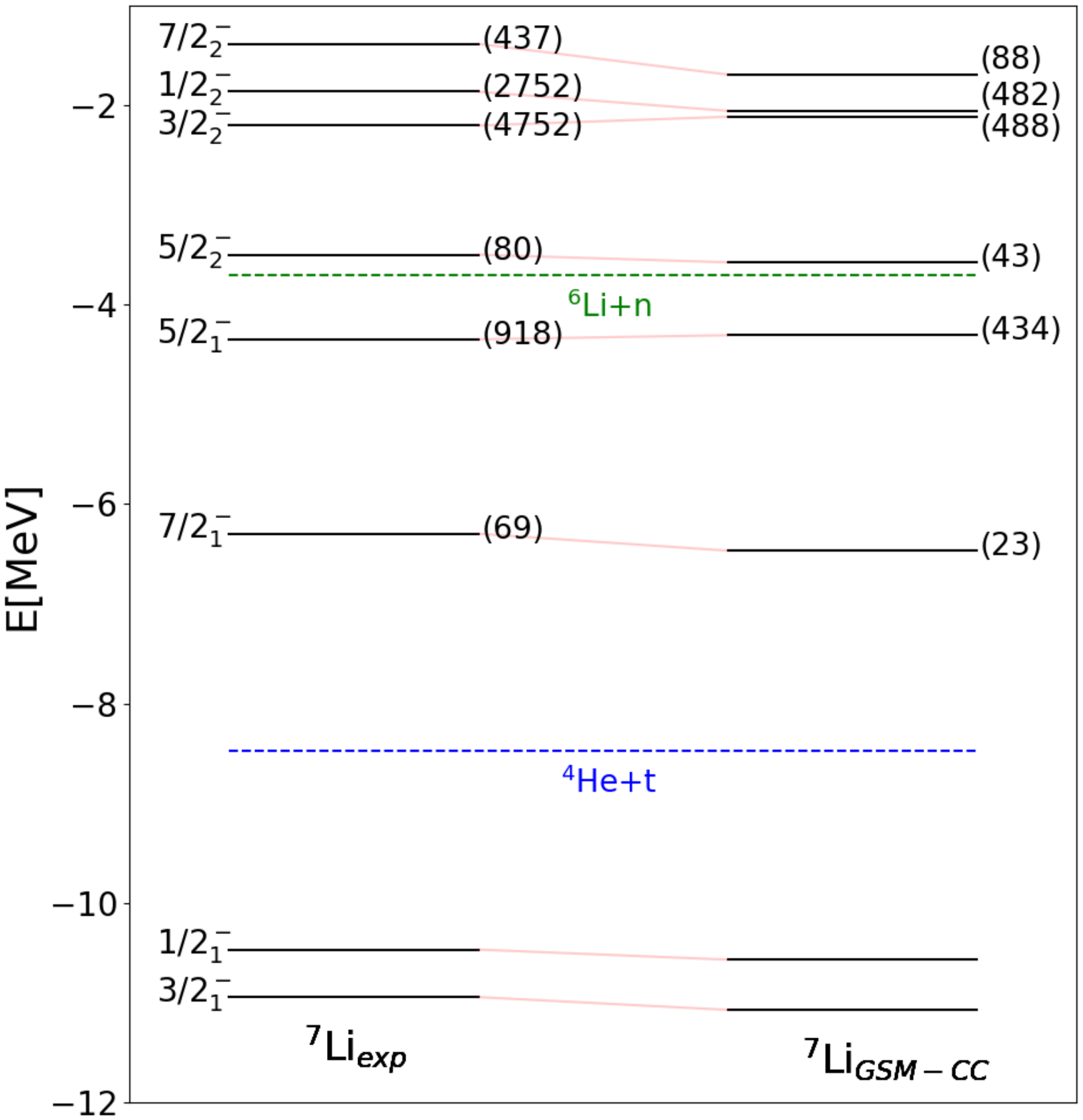}
\end{center}
\vskip -1.5truecm
\caption{GSMCC energy spectrum  of $^7$Li, calculated in the channel basis with two mass partitions:  $[{^4}{\rm He}(0^+_1)\otimes{^3}{\rm H}(L_{j=J})]^{J^{\pi}}$, and $[{^6}{\rm Li}(K^{\pi})\otimes{\rm n}(\ell_j) ]^{J^{\pi}}$, is compared with the data. Numbers in the brackets indicate the resonance width. }
\label{spec_7Li}
\end{figure}

We shall now discuss the spectrum of $^7$Li in the basis comprising $[{^4}{\rm He}(0^+_1)\otimes{^3}{\rm H}(L_j)]^{J^{\pi}}$ and $[{^6}{\rm Li}(K^{\pi})\otimes{\rm n}(\ell_j)]^{J^{\pi}}$ channels. Figure \ref{spec_7Li} shows the GSMCC spectrum of $^7$Li. 
Channels with one-neutron projectile are built by coupling the $^6$Li wave functions with $K^{\pi}$ = $1^+_1$, $1^+_2$, $3^+_1$, $0^+_1$, $2^+_1$, $2^+_2$, with the neutron wave functions in partial waves $\ell_j$: $s_{1/2}$, $p_{1/2}$, $p_{3/2}$, $d_{3/2}$, $d_{5/2}$, $f_{5/2}$, $f_{7/2}$. The cluster channels $[{^4}{\rm He}(0^+_1)\otimes{^3}{\rm H}(L_j)]^{J^{\pi}}$ are constructed by coupling $^3$H wave function in partial waves $L_j$: $s_{1/2}$, $p_{1/2}$, $p_{3/2}$, $d_{3/2}$, $d_{5/2}$, $f_{5/2}$, $f_{7/2}$, with the inert $^4$He core in $K^{\pi} = 0^+$ state.

\begin{figure}[htb]
\vskip -4.5truecm
\begin{center}
\includegraphics[width=11.5cm]{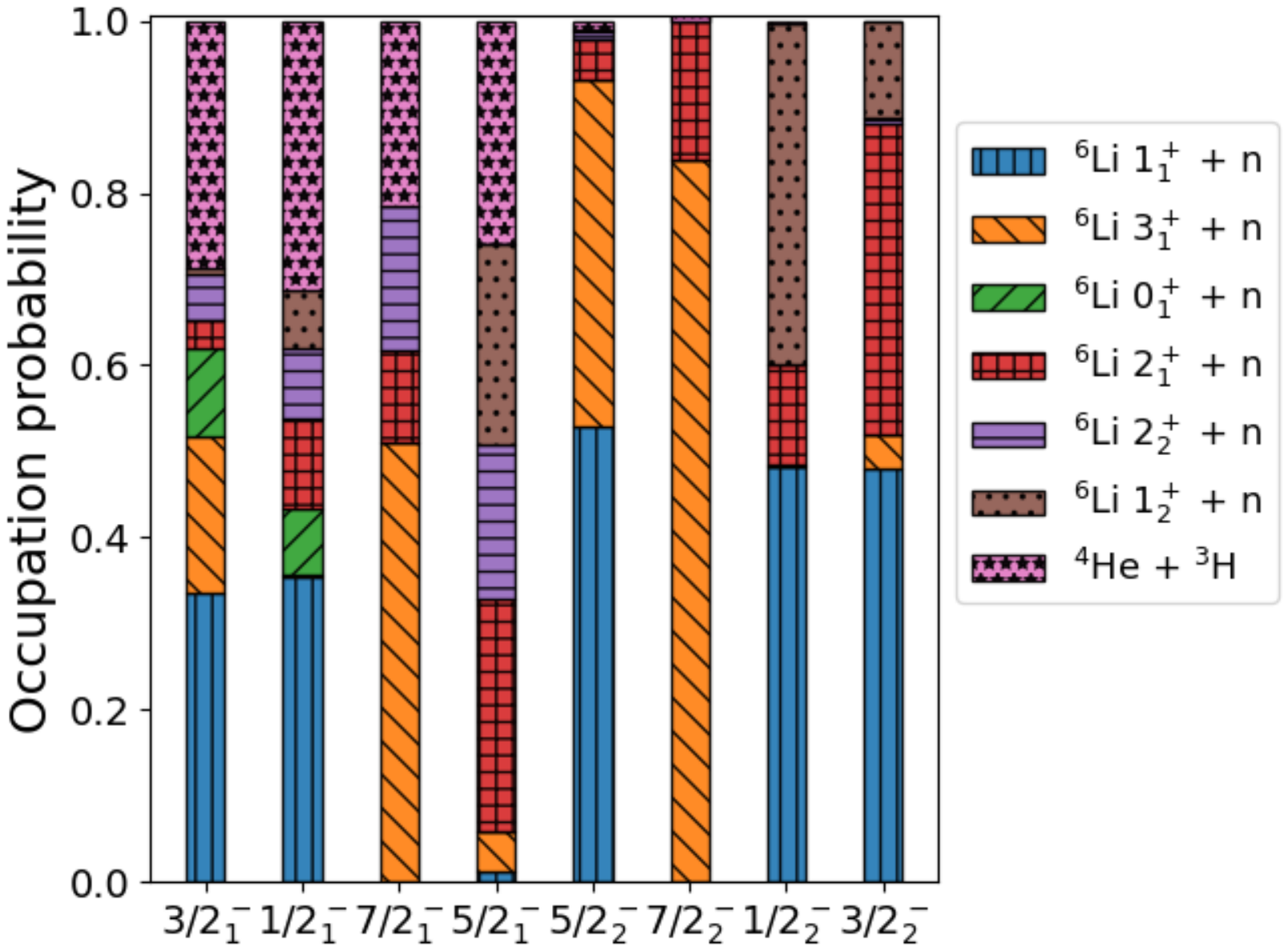}
\end{center}
\vskip -5truecm
\caption{Channel decomposition for selected states  of $^7$Li. In different colors, we show the GSMCC amplitudes of neutron channels  for different angular momenta $K^{\pi}$ of $^6$Li, and the amplitude of the $^4$He + $^3$H cluster channel. Detailed information about major channel amplitudes in the lowest states of $^7$Li and the relevant spectroscopic factors calculated in GSM can be found in Table \ref{table_channels_7Li}. 
 }

\label{occ_7Li}
\end{figure}

Major amplitudes of the channels $[{^4}{\rm He}(0^+_1)\otimes{^3}{\rm H}(L_{j=J})]^{J^{\pi}}$, $[{^6}{\rm Li}(K^{\pi})\otimes{\rm n}(\ell_j) ]^{J^{\pi}}$ are given in Fig. \ref{occ_7Li} and in Table \ref{table_channels_7Li}. A significant probability of the channel wave function $[{^4}{\rm He}(0^+_1)\otimes{^3}{\rm H}(L_{j=J})]^{J^{\pi}}$ is seen only in the low-energy states: $J^{\pi}_i = 3/2^-_1, 1/2^-_1, 7/2^-_1$ which are close to the $^4$He+$^3$H threshold. At higher excitation energies, the probability of the $^4$He+$^3$H channel diminishes rapidly below 1\%.

The ground state $3/2^-_1$ and the first excited state $1/2^-_1$ are dominated by the channels $[{^4}{\rm He}(0^+_1)\otimes{^3}{\rm H}(L_{j=J})]^{J^{\pi}}$ and $[{^6}{\rm Li}(1^+_1)\otimes{\rm n}(p_{1/2}) ]^{J^{\pi}}$. 
In the $7/2^-_1$ resonance, the dominant contribution to the resonance wave function comes from closed neutron channels: $[{^6}{\rm Li}(1^+_1)\otimes{\rm n}(p_{3/2,1/2}) ]^{7/2^-_1}$, $[{^6}{\rm Li}(2^+_2)\otimes{\rm n}(p_{3/2}) ]^{7/2^-_1}$, and the open $^3$H channel: $[{^4}{\rm He}(0^+_1)\otimes{^3}{\rm H}(7/2^-)]^{7/2^-_1}$. 

The $5/2^-_1$ resonance has still a significant component of the open channel $[{^4}{\rm He}(0^+_1)\otimes{^3}{\rm H}(5/2^-)]^{5/2^-_1}$. However, the dominant contribution in the wave function of this resonance comes from the closed neutron channels: $[{^6}{\rm Li}(2^+_1)\otimes{\rm n}(p_{3/2,1/2}) ]^{5/2^-_1}$,  $[{^6}{\rm Li}(1^+_2)\otimes{\rm n}(p_{3/2}) ]^{5/2^-_1}$. The contribution of the open neutron channel $[{^6}{\rm Li}(1^+_1)\otimes{\rm n}(p_{3/2}) ]^{5/2^-_1}$ amounts to $\sim$2\%. 
The close lying $5/2^-_2$ resonance has a completely different structure than the $5/2^-_1$ resonance. 
\begin{table}
	\centering
	\caption{\label{table_channels_7Li} Major GSM-CC amplitudes of channels $[{^6}{\rm Li}(K^{\pi})\otimes{\rm n}\ell_j ]^{J^{\pi}}$ and $[{^4}{\rm He}(0^+_1)\otimes{^3}{\rm H}(K^{\pi})]^{J^{\pi}}$ in $^7$Li. ${\cal R}$[${\tilde c}$] denotes real part of the channel amplitude. ${\cal R}$[${\rm S}$] corresponds to the real part of the GSM spectroscopic factor.}
		\begin{tabular}{cccccc}
			$^{7}$Li ; J$^{\pi}$ & $^{6}$Li ; K$^{\pi}$ & ${^3}{\rm H}$ ; $L_{j}$ & ${\rm n}$ ; $\ell_j$ & ${\cal R}$[${\tilde c}$] & ${\cal R}$[${\rm S}$] \\
			\hline
			${ {1/2}_{1}^{-} }$  & ${ {1}_{1}^{+} }$    &                  &  ${ {p}_{3/2} }$ & 0.33 & 0.80\\
			&                      & ${ ^2{P}_{1/2}}$ &                  & 0.31 & -- \\
			& ${ {2}_{1}^{+} }$    &                  &  ${ {p}_{3/2} }$ & 0.11 & 0.28\\
			& ${ {2}_{2}^{+} }$    &                  &  ${ {p}_{3/2} }$ & 0.08 & 0.21\\
			& ${ {0}_{1}^{+} }$    &                  &  ${ {p}_{1/2} }$ & 0.08 & 0.23\\
			& ${ {1}_{2}^{+} }$    &                  &  ${ {p}_{1/2} }$ & 0.06 & 0.15\\
						\hline
			${ {3/2}_{1}^{-} }$  &                      & ${ ^2{P}_{3/2}}$ &                  & 0.28 & -- \\
			& ${ {1}_{1}^{+} }$    &                  &  ${ {p}_{3/2} }$ & 0.23 & 0.51\\
			& ${ {3}_{1}^{+} }$    &                  &  ${ {p}_{3/2} }$ & 0.18 & 0.50\\
			& ${ {1}_{1}^{+} }$    &                  &  ${ {p}_{1/2} }$ & 0.11 & 0.24\\
			& ${ {0}_{1}^{+} }$    &                  &  ${ {p}_{3/2} }$ & 0.10 & 0.24\\
			& ${ {2}_{1}^{+} }$    &                  &  ${ {p}_{1/2} }$ & 0.03 & 0.09\\
			& ${ {2}_{2}^{+} }$    &                  &  ${ {p}_{3/2} }$ & 0.03 & 0.08\\
			& ${ {2}_{2}^{+} }$    &                  &  ${ {p}_{1/2} }$ & 0.02 & 0.09\\
			\hline
			${ {5/2}_{1}^{-} }$  &                      & ${ ^2{F}_{5/2}}$ &                  & 0.26 & 1.00\\
			& ${ {1}_{2}^{+} }$    &                  &  ${ {p}_{3/2} }$ & 0.23 & 0.53\\
			& ${ {2}_{1}^{+} }$    &                  &  ${ {p}_{3/2} }$ & 0.19 & 0.46\\
			& ${ {2}_{2}^{+} }$    &                  &  ${ {p}_{1/2} }$ & 0.11 & 0.31\\
			& ${ {2}_{1}^{+} }$    &                  &  ${ {p}_{1/2} }$ & 0.08 & 0.16\\
			& ${ {2}_{2}^{+} }$    &                  &  ${ {p}_{3/2} }$ & 0.07 & 0.13\\
			& ${ {3}_{1}^{+} }$    &                  &  ${ {p}_{3/2} }$ & 0.05 & 0.15\\
						\hline
			${ {5/2}_{2}^{-} }$  & ${ {1}_{1}^{+} }$    &                  &  ${ {p}_{3/2} }$ & 0.54 & 0.63\\
			& ${ {3}_{1}^{+} }$    &                  &  ${ {p}_{1/2} }$ & 0.25 & 0.43\\
			& ${ {3}_{1}^{+} }$    &                  &  ${ {p}_{3/2} }$ & 0.15 & 0.25\\
			& ${ {2}_{1}^{+} }$    &                  &  ${ {p}_{3/2} }$ & 0.05 & 0.10\\
			&                      & ${ ^2{F}_{5/2}}$ &                  & 0.01 & 0.12\\
			\hline
			${ {7/2}_{1}^{-} }$  & ${ {3}_{1}^{+} }$    &                  &  ${ {p}_{3/2} }$ & 0.32 & 0.75\\
			&                      & ${ ^2{F}_{7/2}}$ &                  & 0.21 & 1.00\\
			& ${ {3}_{1}^{+} }$    &                  &  ${ {p}_{1/2} }$ & 0.19 & 0.40\\
			& ${ {2}_{2}^{+} }$    &                  &  ${ {p}_{3/2} }$ & 0.17 & 0.42\\
			& ${ {2}_{1}^{+} }$    &                  &  ${ {p}_{3/2} }$ & 0.11 & 0.24\\    
			\hline
		\end{tabular}
	\end{table}
The amplitude of $^3$H channel $[{^4}{\rm He}(0^+_1)\otimes{^3}{\rm H}(5/2^-)]^{5/2^-_2}$ is only $\sim$1\% and the wave function is dominated by the open neutron channel: $[{^6}{\rm Li}(1^+_1)\otimes{\rm n}(p_{3/2}) ]^{5/2^-_2}$, and the closed neutron channels: $[{^6}{\rm Li}(3^+_1)\otimes{\rm n}(p_{3/2,1/2}) ]^{5/2^-_2}$. Hence, we predict that $5/2^-_1$ resonance is excited mainly in $^4$He+$^3$H reaction and decays mostly by the emission of $^3$H, whereas $5/2^-_2$ resonance is excited mainly in $^6$Li+n reaction and decays predominantly by the neutron emission. 

In the $7/2^-_2$ resonance, the summed probability of $[{^6}{\rm Li}(3^+_1)\otimes{\rm n}(p_{3/2,1/2}) ]^{7/2^-_2}$ amounts to 85\% and the weight of the cluster channel $^4$He+$^3$H is totally negligible. 
In the resonances $3/2^-_2$ and $1/2^-_2$, the channels $[{^6}{\rm Li}(1^+_1)\otimes{\rm n}(p_{3/2,1/2}) ]^{J^{\pi}}$ dominate with a summed probability 47\%. Slightly smaller contributions come from the channels $[{^6}{\rm Li}(1^+_2)\otimes{\rm n}(p_{3/2,1/2}) ]^{1/2^-_2}$, $[{^6}{\rm Li}(2^+_1)\otimes{\rm n}(p_{3/2,1/2}) ]^{3/2^-_2}$ for $1/2^-_2$ and $3/2^-_2$, respectively. Other channel wave functions, including the  $^4$He+$^3$H channel, have a negligible weight in these states.

Table \ref{table_channels_7Li} shows also the real part of $^3$H and neutron spectroscopic factors for different states of $^7$Li.
In GSM, the one-nucleon spectroscopic factors~\cite{Michel2007,Michel2007a}, which involve the summation over discrete resonant states and integration along the contour of scattering states of the Berggren ensemble, are independent of the choice of the single-particle basis \cite{Michel2007}. One may notice in Table \ref{table_channels_7Li} a good qualitative agreement between the real part of the one-proton channel amplitude and the corresponding spectroscopic factor, i.e. the large channel amplitudes: $[{^6}{\rm Li}(K^{\pi})\otimes{\rm n}(\ell_j) ]^{J^{\pi}}$, are closely related with the large spectroscopic factors: $^7$Be$(J^{\pi}) \rightarrow {\rm n}(\ell_j) \oplus {^6}{\rm Li}(K^{\pi})$. Hence the channel amplitudes and the spectroscopic factors, both independent of the choice of the single-particle basis provide useful information about the structure of the many-body states. $^3$H spectroscopic factors are calculated by the ratio between the $^3$H channel partial width and the total width. The partial widths are calculated using the current formula \cite{barmore2000theoretical,michel_book_2021}.

The reaction cross-section $^4$He($^3$H, $^3$H)$^4$He is calculated by coupling the real-energy incoming partial waves to the states of $^{6}$Li given by the Hermitian Hamiltonian.
\begin{figure}[htb]
 \vskip -2truecm
\begin{center}
\includegraphics[width=9cm,angle=270]{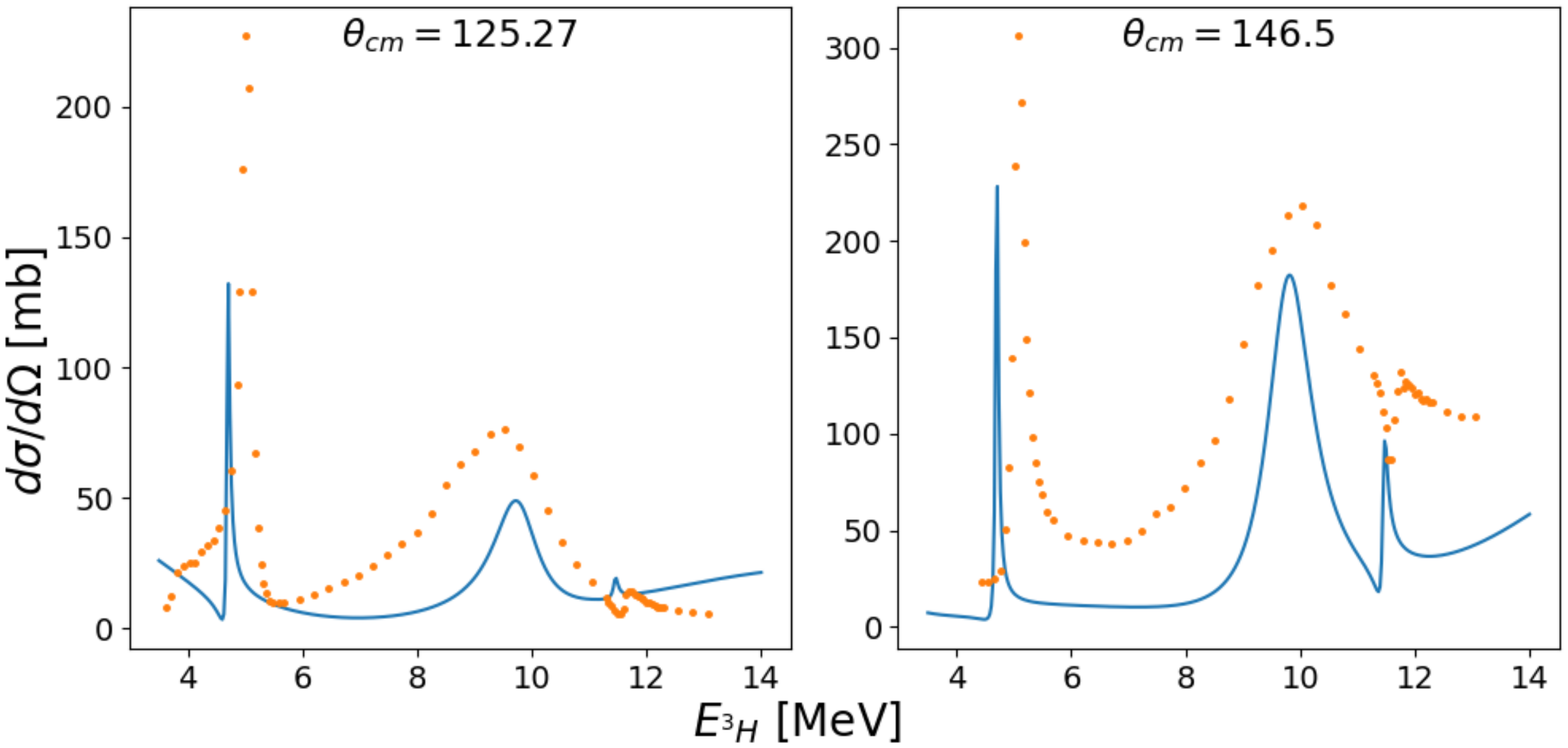}
\end{center}
\vskip -1.8truecm
    \caption{(Color online) 
    The GSMCC differential cross sections of the $^{4}$He($^{3}$H,$^{3}$H)$^{4}$He reaction calculated at two CM angles 
       are compared with the experimental data (in dots) \cite{spiger1967scattering}. }
    \label{fig3}
  \end{figure}
 Fig.  \ref{fig3} shows the $^{4}$He($^{3}$H,$^{3}$H)$^{4}$He cross section calculated in GSMCC (solid line). The calculation is performed with the same Hamiltonian and the same model space as used in the calculation of spectra of $^{6,7}$Li. $^3$H energies in Fig. \ref{fig3} are in the CM
 The estimated error on experimental results varies with the $^3$H bombarding energy and amounts to $\sim$10\% \cite{spiger1967scattering}. The peaks in the calculated cross section correspond to $7/2^-_1$, $5/2^-_1$, and $5/2^-_2$ resonances.

\section{Outlook}
\label{outlook}
Detailed GSMCC analysis provided a microscopic understanding of the imprint of the $^3$H- and 1n-decay channels on the structure of $^7$Li and the $^{4}$He($^{3}$H,$^{3}$H)$^{4}$He reaction cross section. One could see how the $^3$H clusters appear naturally in the many-body states close to the $^4$He+$^3$H reaction channel and disappear in more distant states. This phenomenon, called the near-threshold alignment of many-body states with a nearby reaction channel \cite{okolowicz2012,okolowicz2013}, is expected to play an important role in many astrophysical phenomena changing reaction rates at stellar energies. 

The great advantage of the GSMCC formulation in core+valence particle model space is that it can be applied also in the heavy nuclei where {\it ab initio} approaches are impractical. 

In the future, we plan to extend the unified GSMCC approach for a description of low-energy transfer and radiative captures reactions of astrophysical interest. \\ \\
{\it Acknowledgments}\\
This work has been supported by the National Natural Science Foundation of China under Grant Nos. 12175281 and 11975282; the Strategic Priority Research Program of Chinese Academy of Sciences under Grant No. XDB34000000; the State Key Laboratory of Nuclear Physics and Technology, Peking University under Grant No. NPT2020KFY13.


\bibliographystyle{unsrt}

\bibliography{refs}

\end{document}